\documentclass[sigconf, authorversion]{acmart}
\usepackage{subcaption}
\usepackage{graphicx}

\setcopyright{rightsretained}
\begin{document}
\title{Demo: iJam with Channel Randomization}
\author{Jordan L. Melcher, Yao Zheng}
\orcid{0002-3338-6400}
\author{Dylan Anthony}
\orcid{0000-0000-0000}
\author{Matthew Troglia}
\affiliation{%
}
\author{Thomas Yang, Alvin Yang, Samson Aggelopoulos}
\affiliation{%
  \institution{University of Hawai'i at Mānoa}
  \streetaddress{2540 Dole St}
  \city{Honolulu}
  \state{Hawaii}
  \postcode{96822}
}
\author{Yanjun Pan}
\orcid{0000-0000-0000}
\author{Ming Li}
\orcid{0000-0000-0000}
\affiliation{%
  \institution{University of Arizona}
  \streetaddress{Tucson, AZ 85721}
  \city{Tucson}
  \state{Arizona}
  \postcode{85721}
}

\begin{abstract}
Physical-layer key generation methods utilize the variations of the communication channel to achieve a secure key agreement between two parties with no prior security association. Their secrecy rate (bit generation rate) depends heavily on the randomness of the channel, which may reduce significantly in a stable environment. Existing methods seek to improve the secrecy rate by injecting artificial noise into the channel. Unfortunately, noise injection cannot alter the underlying channel state, which depends on the multipath environment between the transmitter and receiver. Consequently, these methods are known to leak key bits toward multi-antenna eavesdroppers, which is capable of filtering the noise through the differential of multiple signal receptions. This work demonstrates an improved approach to reinforce physical-layer key generation schemes, e.g., channel randomization. The channel randomization approach leverages a reconfigurable antenna to rapidly change the channel state during transmission, and an angle-of-departure (AoD) based channel estimation algorithm to cancel the changing effects for the intended receiver. The combined result is a communication channel stable in the eyes of the intended receiver but randomly changing from the viewpoint of the eavesdropper. We augmented an existing physical-layer key generation protocol, iJam, with the proposed approach and developed a full-fledged remote instrumentation platform to demonstrate its performance. Our evaluations show that augmentation does not affect the bit error rate (BER) of the intended receiver during key establishment but reduces the eavesdropper's BER to the level of random guessing, regardless of the number of antennas it equips. 
\end{abstract}


\copyrightyear{2020}
\acmYear{2020}
\acmConference[WiSec '20]{13th ACM Conference on Security and Privacy in Wireless and Mobile Networks}{July 8--10, 2020}{Linz (Virtual Event), Austria}
\acmBooktitle{13th ACM Conference on Security and Privacy in Wireless and Mobile Networks (WiSec '20), July 8--10, 2020, Linz (Virtual Event), Austria}\acmDOI{10.1145/3395351.3401705}
\acmISBN{978-1-4503-8006-5/20/07}


\maketitle
\section{Introduction}
Physical layer key generation schemes aim to establish a shared key between two parties through an open channel eavesdropped by an adversary. The majority of the schemes leverage the changing wireless channel to generate the key bits, with a generation rate proportional to the entropy of the channel. The more random the channel the faster the key is generated. A few other designs seek to make key generation rate independent from the channel variation, by injecting artificial noise into the wireless channel to aid security \cite{Goel_S,Zhou_X,Liao_W,Li_Q,Goeckel_D}. The combination of the transmission and jamming signal introduces uncertainty to an eavesdropper in the form of noise \cite{Kim_Y,Vilela_JP,Hu_J}, and prevents the eavesdropper from obtaining the underlying key bits.

Gollakota et al. developed iJam, a robust friendly-jamming system which improved the physical-layer key generation for stationary wireless networks \cite{Gollakota_S}. The scheme lets the transmitter (Alice) interleave two identical sequences of bits while the intended receiver (Bob) jams one at random. Since Bob knows which bits are jammed and which ones are clear, he can select the clear bits and reconstruct the key, whereas the eavesdropper (Eve), being unaware of Bob's jamming targets, cannot recreate the key. To increase the randomness of the channel, the key is repetitively transmitted until Bob reconstructs the original message.

Although iJam can achieve secure key exchange in a static channel, Steinmetzer et al. identified a vulnerability by implementing a multi-antenna adversarial model designed to take advantage of the spatial variance to discern between the jammed and clean signals \cite{Steinmetzer_D}. The root of this vulnerability is due to the fact that the channel states between Alice and Eve remain unchanged in spite of Bob's jamming signal, which allows Eve, equipped with multiple antennas, to exploit the pilot or known symbols in Alice's transmission to estimate the channel and cancel its effect. Once Eve equalizes the channel, she may evaluate the signal divergence among multiple antennas to identify the clear symbol. Specifically, a symbol with a large divergence among multiple antennas is likely to be jammed and thus ignored. After a few iterations, clean transmission can be spliced together, and the key can be known.

We propose a defense mechanism against such an attack by combining channel randomization with prediction-based channel equalization. Channel randomization has been used to strengthen physical-layer security schemes, such as orthogonal blinding, that are known to be vulnerable against multi-antenna eavesdropper \cite{Hou_Y,Aono_T,Pan_Y_1,Hassanieh_H}. The method leverages a reconfigurable or moving antenna to create artificial changes in a wireless channel, resulting in unstable channel state information (CSI) between the transmitter and receivers. The prediction-based channel equalization cancels the randomizing effect for Bob by implementing an angle-of-departure (AoD) estimation algorithm to predict the CSI for any given antenna configuration. The combined results are that the channel state appears stable for Bob but continuously changing for Eve. Furthermore, the channel prediction eliminates the need for pilot based channel measurements, which denies Eve the opportunity to measure and cancel the changing effects.

\begin{figure*}[!htb]
\centerline{\includegraphics[width=\linewidth]{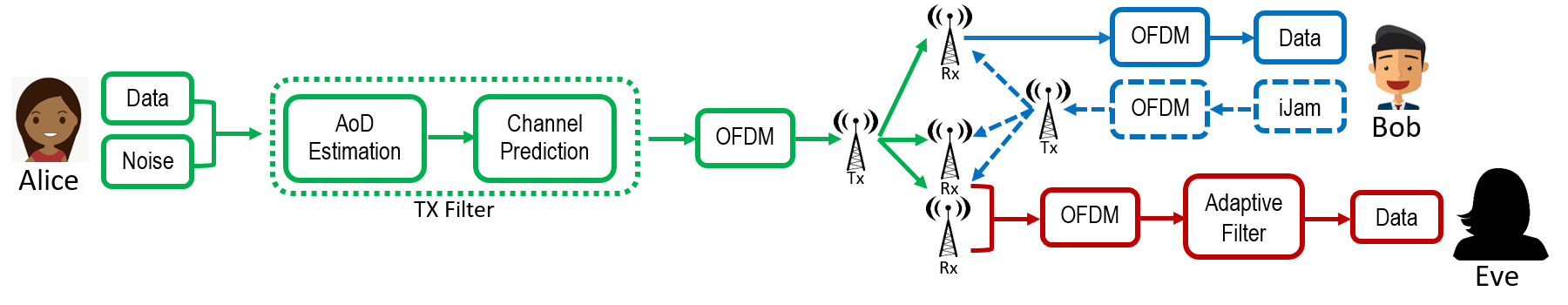}}
\vspace{-15pt}
\caption{System Diagram}
\label{fig1}
\end{figure*}


In this demonstration, we developed and implemented the channel randomizing iJam system with a custom reconfigurable antenna and real-time AoD based channel prediction algorithm, to enhance the security of the key generation protocol. The audience is granted access to our remotely accessible platform via live-stream to control and observe the real-time effects of channel randomization on Bob and Eve. The increase in entropy contributes to both the key generation rate and the overall security. Our results indicate the CSI of the intended receiver bit error rate (BER) does not fluctuate when exposed to channel randomization while simultaneously worsening a multi-antennae adversary's BER to the level of random guessing.

\section{System Overview}
Consider an OFDM system shown in figure \ref{fig1}, where there are three active parties; Alice, who wants to establish a secret key with an intended receiver, Bob, and a passive eavesdropper, Eve. Alice, in green, equips a rotating antenna to randomize the channel, and a digital RF chain consisting of a compressed sensing-based AoD estimation algorithm and a precoding filter to predict and cancel the channel randomization effect for the intended receiver. Bob, in blue, comprises of two stationary antennas, one for receiving, and the other for jamming; while Eve, in red, uses two antennas and an adaptive filter to exploit the spatial variance of the jammed signal. 

\subsection{Channel Randomization}
The channel randomization is implemented by rotating a log-periodic dipole array (LPDA) antenna using a stepper motor. The half-width beam angle of the LDPA antenna's main lobe is approximately 60 degrees. Hence, rotating the antenna by 60 degrees can significantly change the channel state. The rotating speed of the step motor can vary from 1 RPM to 5 RPM. The rotation is carried out in sync with the data transmission, which prevents Eve from obtaining a channel equalization filter to correct the randomization effects.

\subsection{Channel Prediction and Equalization}
In our scheme, Alice predicts and cancels the channel randomization effect for Bob with a compressed sensing-based AoD estimation algorithm. The CSI between Alice and Bob (or Eve) is due to the combined result of multipath components and antenna patterns. The compressed sensing algorithm allows Alice to estimate the 360-degree AoD vector, which defines the multipath components. Given that Alice knows the antenna pattern for a specific antenna mode, the CSI can be computed as the inner product of the AoD distribution and the antenna pattern vector.

To predict the wireless channel, Alice selects every antenna mode then transmits pilot symbols to Bob at each mode. Bob measures the corresponding CSI and sends it back to Alice using implicit feedback. Alice then estimates the AoD vector with the compressed sensing algorithm. Once known, Alice can predict the CSIs for all unused antenna modes and computes the corresponding precoding filter to cancel the changing channel effects for Bob.


\subsection{Key Generation}
With the addition of channel randomization and prediction we created a less complicated physical layer key generation method. In the original scheme, iJam, Alice and Bob switched roles between transmitting and jamming to prevent adversaries who were unaffected by the jamming signals. Channel randomization replaces this method. In our scheme, Alice sends clean OFDM symbols to Bob. Simultaneously, Bob transmits jamming signal at randomly selected time intervals. Alice repeats the same message until Bob has successfully jammed every bit. During this process, Eve compares the captured waveforms from both of her antennas to differentiate between clean and jammed samples.

The bandwidth and frequency are carefully selected to optimize the trade-off between the speed and effect of randomization. The rotating antenna has a limit of 5 RPM with five unique channel modes. To put less strain on the AoD algorithm while providing ample security, 16-QAM is selected. To successfully transmit the repeated key before a new channel mode is selected, the transmission bandwidth is limited to 3.4 KHz. The key bits are randomly generated at Alice's side and shared with Bob through the software detailed in the appendix.

\begin{figure*}[t]
\begin{subfigure}[t]{0.23\linewidth}
  \centering
  \includegraphics[width=\linewidth]{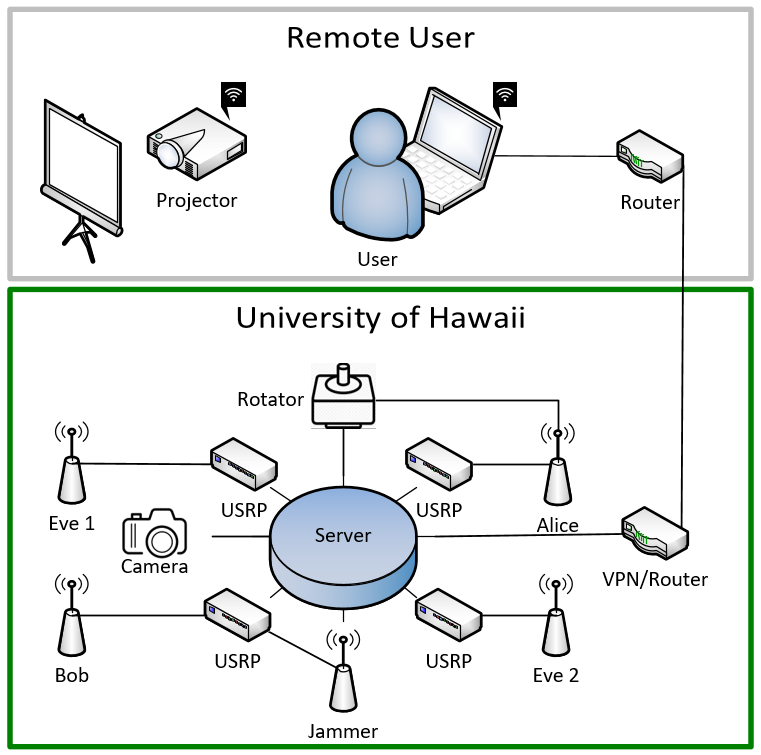}
  \caption{}
\end{subfigure}
\hspace{0.05\textwidth}
\begin{subfigure}[t]{0.23\linewidth}
  \centering
  \includegraphics[width=\linewidth]{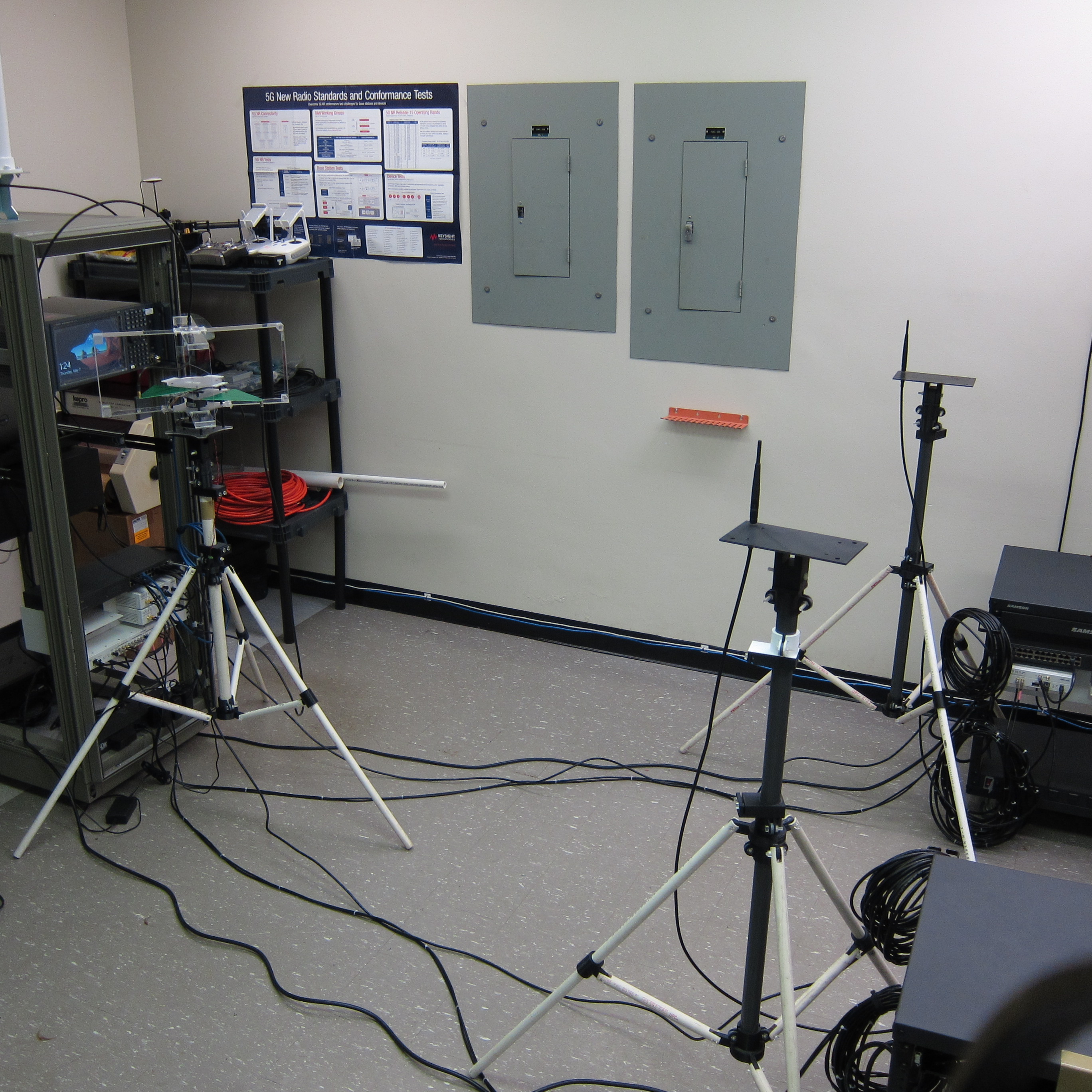}
  \caption{}
\end{subfigure}%
\hspace{0.05\textwidth}
\begin{subfigure}[t]{0.23\linewidth}
  \centering
  \includegraphics[width=\linewidth]{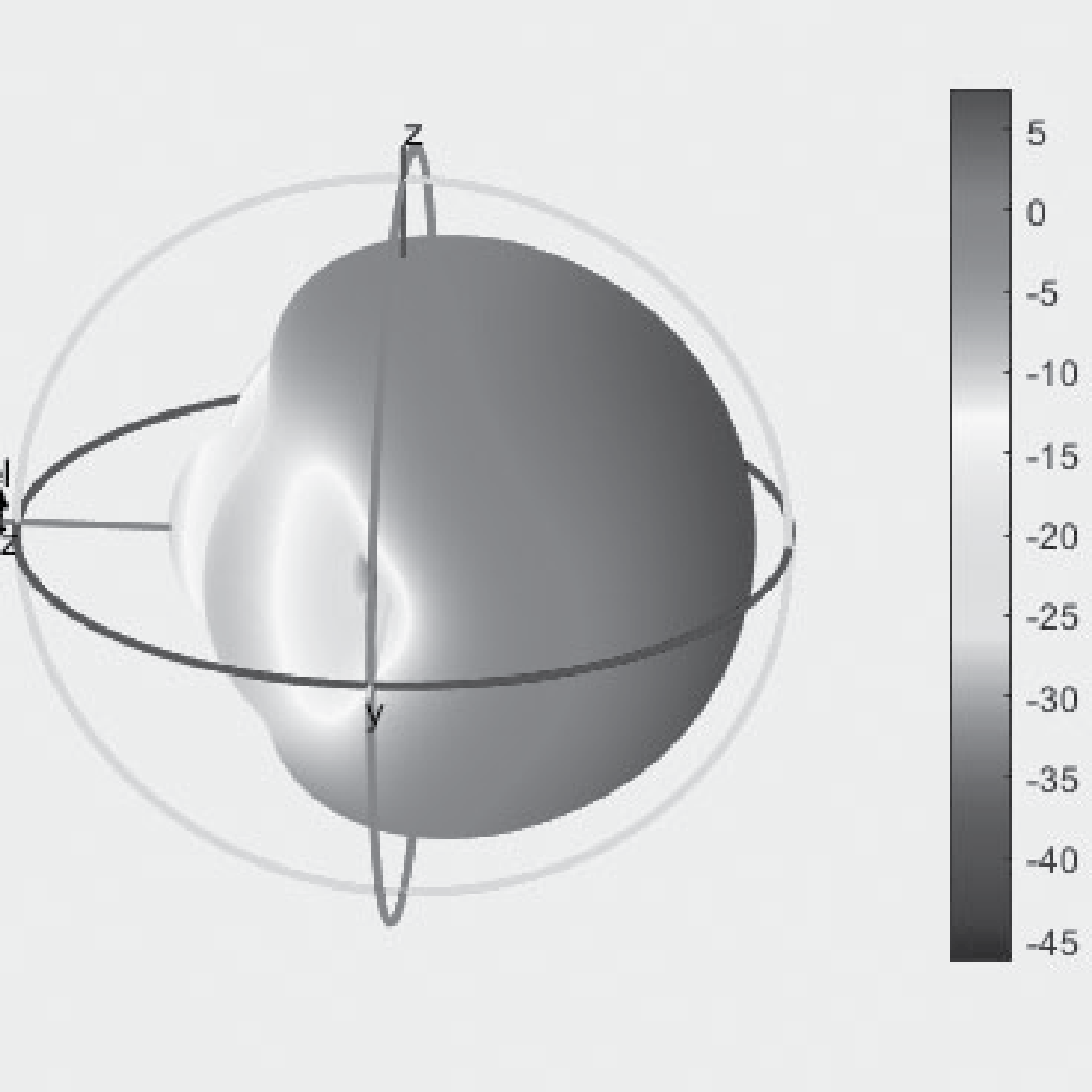}
  \caption{}
\end{subfigure}
\begin{subfigure}[t]{0.23\linewidth}
  \centering
  \includegraphics[width=\linewidth]{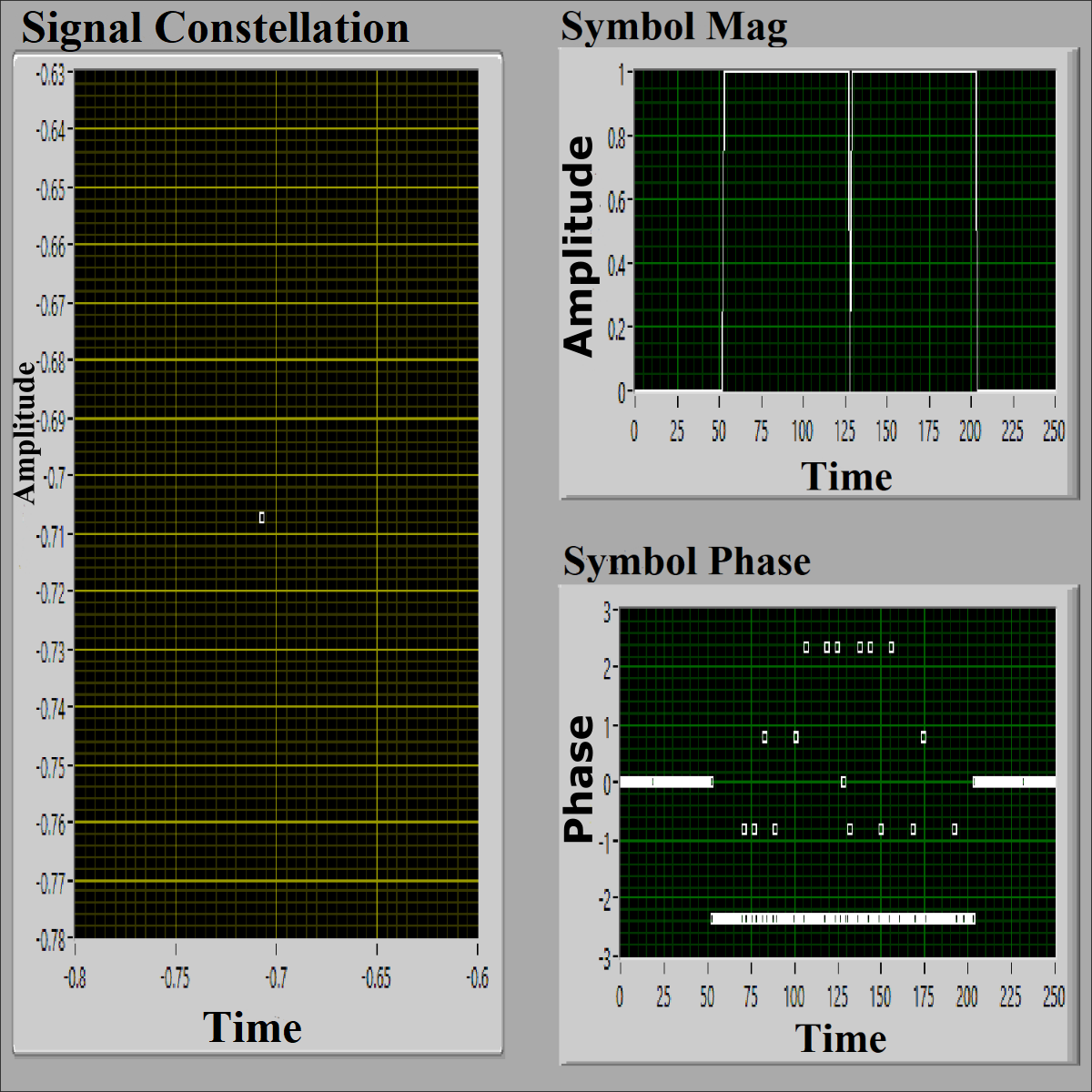}
  \caption{}
\end{subfigure}
\hspace{0.05\textwidth}
\begin{subfigure}[t]{0.23\linewidth}
  \centering
  \includegraphics[width=\linewidth]{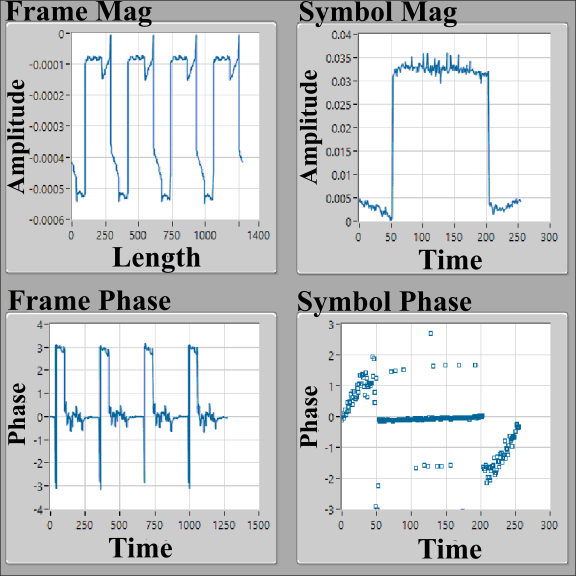}
  \caption{}
\end{subfigure}%
\hspace{0.05\textwidth}
\begin{subfigure}[t]{0.23\linewidth}
  \centering
  \includegraphics[width=\linewidth]{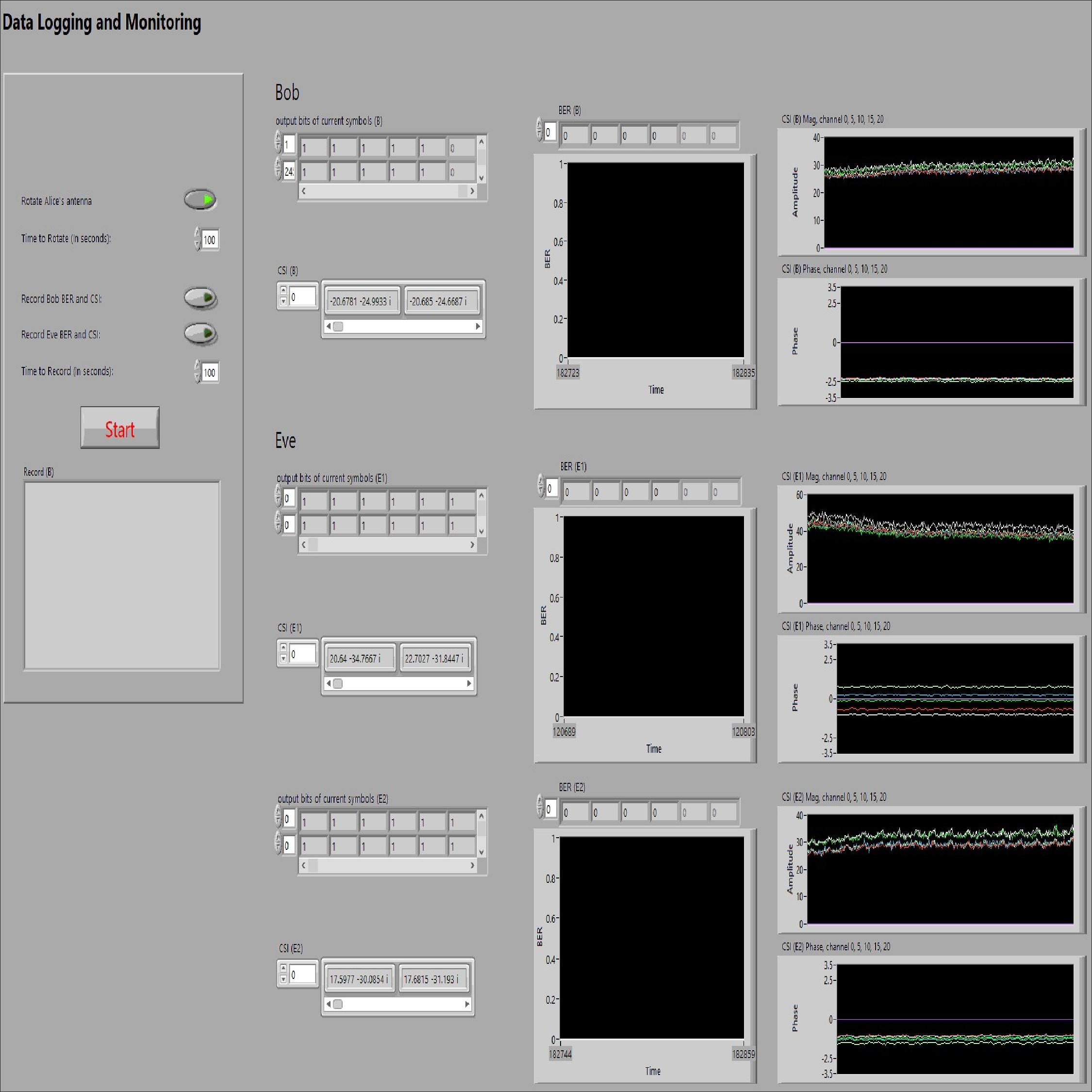}
  \caption{}
\end{subfigure}
\caption{From top to bottom, left to right: (a) Network setup for remote demonstration. (b) Physical setup with Alice on the right, Bob and Eve on the left. (c) The radiation pattern of the LDPA antenna. (d-f) LabView panels to visualize CSIs and data transmissions.}
\label{fig2}
\end{figure*}
\begin{acks}
This work is partly supported by NSF grants CNS-1948568, DGE-1662487, ARO grant W911NF-19-1-0050, and Naval Information Warfare Center Pacific.
\end{acks}
\bibliographystyle{ACM-Reference-Format}
\bibliography{reference}
\appendix
\section{Demonstration Description}
The demonstration platform, consisting of four software-defined radio(s), one-directional LDPA antenna installed on a rotary actuator, three omnidirectional antennas, and one computer server running a customized LabView VI, is hosted in the Cybersecurity Laboratory at the University of Hawai'i at Mānoa (UHM) as displayed in figure \ref{fig2}(b). As illustrated in figure \ref{fig2}(a), remote access to the platform is available through a Virtual Private Network (VPN) gateway implemented on an Amazon EC2 instance. Once connected to the platform, a user can launch the Labview program to rotate the antenna, initiate communications between the USRPs, visualize the wireless signal, and collect the data-traces for offline analysis. A live broadcast will verify the antenna position.

The Labview graphical user interface (GUI), made available to the public, contains the controls and displays necessary for the demonstration. The VI, shown in figure \ref{fig2}(d-f), allows users to perform experiments and analysis. Users can control IQ rate, transmission frequency and gain, bit number generation, rotation rate, and experiment duration. The graphs in figure \ref{fig2}(d), depict the transmitter signal constellation, symbol magnitude, and phase. Figure \ref{fig2}(e) displays magnitude and phase plots for signal reception verification of one of the three receivers. Most importantly, in figure \ref{fig2}(f), Bob and one of Eve's antennas compare their CSI. Two scenarios will run with and without channel randomization to observe the drastic effect on an eavesdropping attack.

\end{document}